\documentclass[multphys,vecphys]{svmult}

\usepackage{makeidx}         
\usepackage{graphicx}        
\usepackage{multicol}        
\usepackage[bottom]{footmisc}

\makeindex             

\begin{document}

\title*{Star Cluster Evolution: From young massive star clusters to
old globulars} 
\titlerunning{From young massive star clusters to old globulars}
\author{Richard de Grijs}
\institute{Department of Physics \& Astronomy, The University of
Sheffield, Hicks Building, Hounsfield Road, Sheffield S3 7RH, UK;
\texttt{R.deGrijs@sheffield.ac.uk}}
%
%
\maketitle

\begin{abstract}
Young, massive star clusters are the most notable and significant end
products of violent star-forming episodes triggered by galaxy
collisions, mergers, and close encounters. The question remains,
however, whether or not at least a fraction of the compact YMCs seen
in abundance in extragalactic starbursts, are potentially the
progenitors of globular cluster (GC)-type objects. However, because of
the lack of a statistically significant sample of similar nearby
objects we need to resort to either statistical arguments or to the
painstaking approach of case by case studies of individual objects in
more distant galaxies. Despite the difficulties inherent to addressing
this issue conclusively, an ever increasing body of observational
evidence lends support to the scenario that GCs, which were once
thought to be the oldest building blocks of galaxies, are still
forming today.
\end{abstract}

\section{Young massive star clusters as proto-globular clusters}

The production of luminous, massive yet compact star clusters (YMCs;
often with masses $m_{\rm cl} \ge 10^5 {\rm M}_\odot$) seems to be a
hallmark of the most intense starbursts. YMCs are therefore important
as benchmarks of cluster formation and evolution. They are also
important as tracers of the (stellar) initial mass function (IMF) and
other physical characteristics in starbursts. The key properties of
YMCs have been explored in starburst regions in several dozen
galaxies, both in normal spirals and in gravitationally interacting
galaxies.

The question remains, however, whether or not at least some of the
YMCs observed in extragalactic starbursts might survive for a Hubble
time. If we could settle this issue convincingly, one way or the
other, the implications would be far-reaching for a wide range of
astrophysical questions, including (but not limited to) our
understanding of the process of galaxy formation, assembly and
evolution, and the process and conditions required for star and star
cluster formation. 

\section{Resolution of the evolutionary question}

The evolution of young clusters depends crucially on their stellar
IMF: if the IMF is too shallow, i.e., if the clusters are
significantly depleted in low-mass stars compared to, e.g., the solar
neighbourhood, they will likely disperse within about a Gyr of their
formation (e.g., Gnedin \& Ostriker 1997; Goodwin 1997; Smith \&
Gallagher 2001; Mengel et al. 2002). At present, there are two
principal approaches in which one can attempt to address the
underlying IMFs of extragalactic YMCs [but see also de Grijs et
al. (2005a) for an alternative approach].

\subsection{The Cluster Luminosity Function: the case of M82}

In de Grijs et al. (2003a,b) we reported the discovery of an
approximately log-normal cluster luminosity and mass function (CLF,
CMF) for the roughly coeval star clusters at the intermediate age of
$\sim 1$ Gyr in M82's fossil starburst region ``B''.  This provided
the first deep CLF (CMF) for a star cluster population at intermediate
age, which thus serves as an important benchmark for theories of the
evolution of star cluster systems [see also Goudfrooij et al. (2004)
for a related important result for NGC 1316, at $\sim 3$ Gyr].

A substantial series of papers on the young Large Magellanic Cloud
cluster system (with ages $\le 2 \times 10^9$ yr), starting with the
seminal work by Elson \& Fall (1985), seem to imply that the CLF of
YMCs is well described by a power law [but see de Grijs \& Anders
(2006) for caveats]. On the other hand, for old globular cluster (GC)
systems with ages $\ge 10^{10}$ yr, the CLF shape is well established
to be roughly log-normal, and almost universal among local galaxies.
This type of observational evidence has led to the popular, but thus
far mostly speculative theoretical prediction that not only a
power-law, but {\it any} initial CLF (CMF) will be rapidly transformed
into a log-normal distribution because of {\it (i)} stellar
evolutionary fading of the lowest-luminosity (mass) objects to below
the detection limit; and {\it (ii)} disruption of the low-mass
clusters due both to interactions with the gravitational field of the
host galaxy, and to internal two-body relaxation effects leading to
enhanced cluster evaporation.

From our detailed analysis of the expected evolution of CMFs starting
from initial log-normal and initial power-law distributions (de Grijs
et al. 2005b), we conclude that our observations of the M82 B CMF are
inconsistent with a scenario in which the 1 Gyr-old cluster population
originated from an initial power-law mass distribution. This applies
to a large range of ``characteristic'' cluster disruption
time-scales. Our conclusion is supported by arguments related to the
initial density in M82 B, which would be unphysically high if the
present cluster population were the remains of an initial power-law
distribution (particularly in view of the effects of cluster ``infant
mortality'', which require large excesses of low-mass unbound clusters
to be present at the earliest times).

In de Grijs et al. (2003c) we showed that the CMFs of YMCs in many
different environments are well approximated by power laws with slopes
$\alpha \simeq -2$. However, except for the intermediate-age cluster
systems in M82 B and NGC 1316 (Goudfrooij et al. 2004), the {\it
expected} turn-over (or peak) mass (based on comparisons with
present-day GC systems and taking evolutionary fading into account) in
most YMC systems observed to date occurs close to or below the
observational detection limit, simply because of their greater
distances and shallower observations. As such, these results are not
necessarily at odds with each other, but merely hindered by
observational selection effects.

\subsection{High-resolution spectroscopy: individual cluster analysis}

With the ever increasing number of large-aperture ground-based
telescopes equipped with state-of-the-art high-resolution
spectrographs and the wealth of observational data provided by the
{\sl Hubble Space Telescope}, we may now finally be getting close to
resolving the potentially far-reaching issue of YMC-to-GC evolution
conclusively. To do so, one needs to obtain {\it (i)} high-resolution
spectroscopy, in order to obtain dynamical mass estimates, and {\it
(ii)} high-resolution imaging to measure their sizes (and
luminosities). As a simple first approach, one could then construct
diagnostic diagrams of YMC mass-to-light ratio vs. age, and compare
the YMC locations in this diagram with models of ``simple stellar
populations'' (SSPs) using a variety of IMF descriptions (cf. Smith \&
Gallagher 2001; Mengel et al. 2002; Bastian et al. 2006). However,
such an approach, while instructive, has serious shortcomings:

{\it (i)} In this simple approach, the data can be described by {\it
both} variations in the IMF slope {\it and} variations in a possible
low-mass cut-off; the models are fundamentally degenerate for these
parameters.

{\it (ii)} While the assumption that these objects are approximately
in virial equilibrium is probably justified at ages greater than a few
$\times 10^7$ yr (at least for the stars dominating the light), the
{\it central} velocity dispersion (as derived from luminosity-weighted
high-resolution spectroscopy) does not necessarily represent a YMC's
total mass. It is now well-established that almost every YMC exhibits
significant mass segregation from very early on, so that the effects
of mass segregation must be taken into account when converting central
velocity dispersions into dynamical mass estimates (see also Lamers et
al. 2006; J.J.  Fleck et al., in prep.).

{\it (iii)} With the exception of a few studies (e.g., M82-F; Smith \&
Gallagher 2001), the majority of YMCs thus far analysed in this way
have ages around 10 Myr. Around this age, however, red supergiants
(RSGs) appear in realistic stellar populations. Unfortunately, the
model descriptions of the RSG phase differ significantly among the
various leading groups producing theoretical stellar population
synthesis codes (Padova vs. Geneva vs. Yale), and therefore the
uncertainties in the evolutionary tracks are substantial.

\section{The current verdict?}

It may appear that a fair fraction of the $\sim 10$ Myr-old YMCs that
have been analysed thus far may be characterised by unusual IMFs,
since their loci in the diagnostic diagram are far removed from any of
the ``standard'' SSP models (see, e.g., Bastian et al. 2006). However,
Bastian \& Goodwin (2006) recently showed that this is most likely an
effect of the fact that the velocity dispersions of these young
objects do not adequately trace their masses. They are instead
strongly affected by the effects of gas expulsion due to supernova
activity and massive stellar winds. In this respect, it is encouraging
to see that the older clusters (i.e., older than M82-F, a few $\times
10^7$ yr) seem to conform to ``normal'' IMFs; by those ages, the
clusters' velocity dispersions seem to represent the underlying
gravitational potential much more closely.

We recently reported the discovery of a extremely massive, but old
($12.4 \pm 3.2$ Gyr) GC in M31, 037-B327, that has all the
characteristics of having been an exemplary YMC at earlier times (Ma
et al. 2006). In order to have survived for a Hubble time, we conclude
that its stellar IMF cannot have been top-heavy, i.e., characterized
by a low-mass cut-off at $m_\star \ge 1$ M$_\odot$, as sometimes
advocated for current YMCs (e.g., Smith \& Gallagher 2001). Using this
constraint, and a variety of SSP models, we determine a photometric
mass for 037-B327 of $M_{\rm GC} = (3.0 \pm 0.5)\times 10^7$
M$_\odot$, somewhat depending on the SSP models used, the metallicity
and age adopted and the IMF representation. In view of the large
number of free parameters, the uncertainty in our photometric mass
estimate is surprisingly small. This mass, and its relatively small
uncertainties, make this object the most massive star cluster of any
age in the Local Group. As a surviving ``super'' star cluster, this
object is of prime importance for theories aimed at describing massive
star cluster evolution.

\printindex
\end{document}